\documentclass[english,aps,prc,twocolumn,showpacs]{revtex4}
\usepackage[T1]{fontenc}
\usepackage[latin1]{inputenc}
\usepackage{graphicx}
\usepackage{babel}
\newcommand{\ba}{\begin{eqnarray}}
\newcommand{\ea}{\end{eqnarray}}

\begin{document}

\title{Eigenvalue correlations and the distribution of ground state 
angular momenta for random many-body quantum systems}

\author{J. Barea}

\affiliation{Center for Theoretical Physics, Sloane Physics Laboratory, 
Yale University, P.O. Box 208210, New Haven, Connecticut 06520-8120, U.S.A. 
\footnote{Present address: Facultad de F{\'{\i}}sica, Universidad de Sevilla, 
Avda. Reina Mercedes s/n, E-4012 Sevilla, Spain}}

\author{R. Bijker and A. Frank}

\affiliation{Instituto de Ciencias Nucleares, 
Universidad Nacional Aut\'onoma de M\'exico, 
A.P. 70-543, 04510 M\'exico, D.F., M\'exico}

\date{\today}

\begin{abstract} 
The observed preponderance of ground states with angular momentum $L=0$ in 
many-body quantum systems with random two-body interactions is analyzed in 
terms of correlation coefficients (covariances) among different eigenstates. 
It is shown that the geometric analysis of Chau {\it et al.} can be interpreted 
in terms of correlations (covariances) between energy eigenvalues thus providing 
an entirely statistical explanation of the distribution of ground state 
angular momenta of randomly interacting quantum systems which, in principle, is 
valid for both fermionic and bosonic systems. The method is illustrated for the 
interacting boson model.  
\end{abstract}

\pacs{05.30.-d, 24.60.Lz, 21.60.-n}

\maketitle

\section{Introduction}

Low-lying spectra of many-body quantum systems often display a high degree 
of order and regularity. In the case of atomic nuclei, despite their 
complexity and the large number of degrees of freedom involved, they often 
exhibit simple features, such as pairing properties, surface vibrations and 
rotational motion in even-even nuclei. Conventional wisdom is that regularities 
arise from symmetries of the Hamiltonian, which lead to invariances that 
severely constrain the many-body motion. While some of these symmetries are exact 
({\it e.g.} rotational and reflection invariance), others are approximate 
({\it e.g.} isospin). These global symmetries, however, do not explain by 
themselves the regular patterns observed. Further assumptions about the 
nature of the nucleon-nucleon interaction are required. Thus, a strongly 
attractive pairing force between like nucleons has been shown to be responsible 
for the remarkable constancy of the excitation energy of the first excited $L=2$ 
states in the Sn isotopes, while deformation and rotational behavior is known to 
arise from an attractive quadrupole-quadrupole interaction between neutrons and
protons \cite{Talmi}. These striking patterns as well as many other correlations
have been shown to be robust features of low-energy nuclear behavior,
which signal the emergence of order and collectivity. In every case
the patterns arise as a consequence of particular forms of the nucleon-nucleon
interaction. Most features of low-lying nuclear spectra have thus
been explained in terms of a short-range pairing interaction and 
a long-range quadrupole force. 

Hence, it came as a big surprise when Johnson, Bertsch and Dean \cite{JBD} found 
that for even-even nuclei shell-model Hamiltonians with random two-body interactions 
(the so-called Two-Body Random Ensemble or TBRE) are very likely to yield a ground 
state with angular momentum $L=0$. In fact, the probability for a $L=0$ state to 
become the ground state turns out to be much larger than that expected on the basis of 
the fraction of $L=0$ states in the model space. Similar regularities were later 
found to exist as well in bosonic \cite{BF} and electronic \cite{electronic} many-body 
quantum systems with random interactions, so the occurrence of these regular 
phenomenona in spite of the random nature of the two-body interactions seems to be 
a rather generic feature. The unexpected results of Johnson {\it et al.} 
are reminiscent of other statistical results associated with correlated sets of numbers, 
which are perceived as contrary to expectations. This is the case, for example, for 
Benford's Law, which deals with the counterintuitive relative frequency distributions 
of digits in a given data set and which is related to scale invariance \cite{benford}.

The observation of a statistical preference of $L=0$ ground states for ensembles 
of random two-body interactions has sparked a large number of investigations to 
further explore the properties of these random systems and to understand the 
mechanism for the emergence of regular ordered spectral features from random 
interactions \cite{NPN,msu,zhao,thomas}. The appearance of ordered spectra in systems 
with chaotic dynamics is a robust property, that does not depend on the 
specific choice of the (two-body) ensemble of random interactions 
\cite{JBD,BFP,DD,ymzhao},  
time-reversal symmetry \cite{BFP}, the restriction of the Hamiltonian to one- 
and two-body interactions \cite{BF2}, nor is it limited to yrast states with 
small angular momentum $L=0,2,4$ \cite{johnson} as used in the original studies 
\cite{JBD,BF}. Despite the progress and new insights in understanding the appearance 
of ordered spectra from random interactions, the proposed explanations offer  
partial solutions of the problem, without being able to account for all observed 
regular phenomena in systems of randomly interacting fermions and bosons. Among others, 
we mention induced pairing \cite{JBDT}, geometric chaoticity for randomly interacting 
fermions \cite{msu,mulhall}, mean-field analysis for the interacting boson model and 
the vibron model \cite{mf}, spectral widths \cite{BFP,papenbrock} and an empirical 
method based on the eigenvalues of each independent two-body matrix element 
\cite{ymzhao}. 

The robustness of the numerical results for both systems of randomly interacting 
fermions and bosons strongly suggests that an explanation of the origin of the 
observed regular features has to be sought in the many-body dynamics of the model 
space and/or the general statistical properties of random interactions, a conclusion 
which is suggested by many theoretical studies. In this respect, the approach of 
Chau {\it et al.} \cite{chau} for diagonal Hamiltonians stands out, since it makes 
use of the geometry of the model space, is valid for both fermions and bosons, 
and allows to calculate the ground state probabilities exactly.  

In this article, we show that the geometric analysis of \cite{chau} 
can be interpretated in terms of correlations (covariances) between energy eigenvalues 
to provide an entirely statistical explanation of the distribution of ground state 
angular momenta of randomly interacting many-body quantum systems. In addition, we 
show that the method can be extended to non-diagonal Hamiltonians by using perturbation 
theory.  

\section{Covariances}

In \cite{chau}, spectroscopic properties of quantum systems with random interactions 
were given a geometric interpretation. In particular, it was shown that diagonal Hamiltonians, 
{\it i.e.} whose energy eigenvalues depend linearly on the two-body matrix elements, can be 
associated with a geometric shape (convex polyhedron) defined in terms of coefficients of 
fractional parentage and/or generalized coupling coefficients. The probability for a certain 
state to become the ground state is then related to the angles at the vertices. 
In this approach, geometry arises as a consequence of strong correlations 
implicit in many-body quantum systems. Random tests can be understood in this context as 
sampling experiments on this geometry. 

Let us consider many-body quantum systems for which the energy eigenvalues $e_i$ can be expressed 
as linear combinations of two-body matrix elements $r_m$ (as is the case for any interaction 
between identical fermions with $j \leq \frac{7}{2}$ or identical bosons with $l \leq 3$) 
\ba
e_{i} = \sum_{m=1}^{d} c_{m}^{i} r_{m} ~.
\ea
The coefficients $c_{m}^{i}$ contain the information on the many-body quantum system 
via angular momentum coupling coefficients, coefficients of fractional parentage, etc. 
The statistical properties of such a system can be studied by taking an ensemble of random  
two-body matrix elements $r_m$ in which the random variables $r_m$ are chosen independently 
on a Gaussian distribution with zero mean and unit width. The covariance coefficients between 
two energies are given by
\begin{eqnarray}
\langle e_{i}e_{j} \rangle &=& \frac{1}{N-1} \sum_{p=1}^{N} 
\bigl( e_{i}^{(p)} - \langle e_{i} \rangle \bigr)
\bigl( e_{j}^{(p)} - \langle e_{j} \rangle \bigr)
\nonumber\\
&=& \sum_{m=1}^{d} c_{m}^{i} c_{m}^{j} = \vec{c}^{\, i} \cdot \vec{c}^{\, j} ~,
\end{eqnarray}
where $N$ is the size of the ensemble and $p$ indicates the $p$-th realization of the 
ensemble. The covariances depend on the relative angle between the vectors 
$\vec{c}^{\, i}$ and $\vec{c}^{\, j}$. 

It was shown by Chau {\em et al.}, that all energies are confined to a convex polytope 
({\em i.e.} a convex polyhedron in $d$ dimensions), and that only the states that are 
located at the vertices of this polyhedron can become the ground state \cite{chau}. 
The probability for a state at vertex $j$ to become the ground state depends on the 
angle $\sum_{f \ni j}\theta_{jf}$, where the sum is over all faces that contain the 
vertex $j$ and $\theta_{jf}$ is the angle subtended at vertex $j$ in the face $f$. 
For $d=2$ and $d=3$ dimensions the explicit forms are
\ba
P_{j}^{(2)} &=& \frac{1}{2}-\frac{\theta_{j}}{2\pi} ~,
\nonumber\\
P_{j}^{(3)} &=& \frac{1}{2}-\frac{1}{4\pi}\sum_{f \ni j}\theta_{jf} ~.
\ea

From these considerations it follows directly that the probability that a certain state 
can become the ground state is related to the covariances of the energies. Suppose that 
the angle $\theta_{jf}$ at vertex $j$ in the face $f$ is defined by the vertices $i$, 
$j$ and $k$ of the convex polygon. A straightforward application of the cosine rule 
shows that 
\begin{equation}
\cos\theta_{jf} = \frac{\langle e_{j}^{2} \rangle  + \langle e_{i}e_{k} \rangle 
- \langle e_{i}e_{j} \rangle - \langle e_{j}e_{k} \rangle} 
{\sqrt{\langle e_{i}^{2} \rangle + \langle e_{j}^{2} \rangle -2 \langle e_{i}e_{j} \rangle}
 \sqrt{\langle e_{k}^{2} \rangle + \langle e_{j}^{2} \rangle -2 \langle e_{j}e_{k} \rangle}} ~.
\label{cov}
\end{equation}
We thus find that the probability for a state which is located at a vertex $j$ of 
the convex polytope to become the ground state is determined by the covariance 
coefficients of the energy eigenvalues. It is important to note that this relation 
is exact and may be used to explain the approximate results based on the spectral widths 
\cite{BFP,papenbrock}. Eq.~(\ref{cov}) is valid for any many-body quantum system whose 
energies are linear functions of the random variables and holds for both bosons and fermions.  

\section{Results}

The relation between the distribution of ground state angular momenta and the 
covariances between energy eigenvalues of Eq.~(\ref{cov}) provides a completely 
statistical interpretation of the distribution of ground state angular momenta 
of randomly interacting many-body quantum systems. To the best of our knowledge, 
this is the first time such a connection has been derived in explicit form. 
Since the geometric method of Chau {\it et al.} is valid for diagonal interactions 
for which the energy eigenvalues depend linearly on the random interactions, 
Eq.~(\ref{cov}) provides a statistical interpretation of the exact results of 
\cite{chau}.  
 
For systems in which one has both diagonal and off-diagonal matrix elements 
the method of covariance coefficients cannot be applied directly, since the 
energy eigenvalues show a nonlinear dependence on the random variables. 
In this section, we show that also for non-diagonal systems the distribution of 
ground state angular momenta can be obtained by treating the off-diagonal 
interactions in perturbation theory. 
In this way, one obtains an approximate expression of the energy eigenvalues 
which is linear in the random variables and hence is amenable to a statistical 
analysis based on the covariances. Obviously, this is only valid as far as the 
distribution of ground state angular momenta is concerned. Realistic ground 
state wave functions have a far more complicated structure that goes way 
beyond the use perturbation theory. 
 
As an illustration, we discuss two schematic Hamiltonians of the Interacting 
Boson Model (IBM), a model for collective excitations in medium and heavy mass 
nuclei \cite{ibm}. 

As a first example, we consider the IBM Hamiltonian
\ba
H_1 = \cos \chi \, d^{\dagger} \cdot \tilde{d} +
\frac{\sin \chi}{4(N-1)} 
( s^{\dagger} s^{\dagger} - d^{\dagger} \cdot d^{\dagger} ) 
( {\rm h.c.} )  ~,
\label{hibm1}
\ea
which describes a transition between vibrational ($\sin \chi=0$) and $\gamma$-unstable 
nuclei ($\cos \chi=0$). In the general case, the Hamiltonian has to diagonalized 
numerically to obtain its eigenvalues. The eigenstates can be labeled by the total 
number of bosons $N$, the boson seniority $v$ and the angular momentum $l$. The ground 
state has either $v=0$, $v=1$ or $v=N$. In the present study the angle $\chi$ is taken 
on the interval $-\pi < \chi \le \pi$, so that all possible combinations of attractive 
and repulsive interactions are covered. For this reason, the Hamiltonian effectively 
depends on two random coefficients $\cos \chi$ and $\sin \chi$ which can be either 
attractive or repulsive.
 
The distribution of ground state angular momenta was obtained exactly in a mean-field 
study \cite{mf}. For even values of the number of bosons $N$, the ground state has 
$v=0$ in 75 \% of the cases and $v=N$ in the remaining 25 \%. For odd values of $N$, 
the ground state is either $v=0$, $v=1$ or $v=N$ for 50 \%, 25 \% and 25 \% of the 
cases, respectively. 
 
The Hamiltonian of Eq.~(\ref{hibm1}) has a spherical ground state for 
$\sin \chi \leq |\cos \chi|$ and a deformed one for $\sin \chi \geq |\cos \chi|$ 
\cite{DSI,mf}. 
In the first case, the appropriate basis is that of the $U(5)$ limit of the IBM, 
$| N,n,v \rangle$. The first term in the Hamiltonian of Eq.~(\ref{hibm1}) is 
diagonal in this basis, whereas the second term contains both a diagonal and a 
nondiagonal contribution. The latter interaction can be treated in first order 
perturbation theory so that the approximate energy spectrum has a linear dependence 
on the parameters
\ba
E_{nv} &=& \cos \chi \, n_d + \frac{\sin \chi}{4(N-1)} [ (N-n)(N-n-1)
\nonumber\\
&& \hspace{2cm} + (n-v)(n+v+3) ] ~.
\label{esph}
\ea
For the deformed solution, the appropriate basis is that of the $SO(6)$ limit of the 
IBM, $|N,\sigma,v \rangle$. In this case, the second term of Eq.~(\ref{hibm1}) is 
diagonal, whereas the first term contains diagonal and off-diagonal contributions. 
In first order perturbation theory, the energy spectrum is given by 
\ba
E_{\sigma v} = \sin \chi \frac{(N-\sigma)(N+\sigma+4)}{4(N-1)}  
+ \cos \chi \sum_n n \left( \zeta_{n \sigma}^v \right)^2 ~,
\label{edef}
\ea
where $\zeta_{n \sigma}^v = \langle N,n,v | N,\sigma,v \rangle$ \cite{mex}.  

In Table~\ref{sphdef1}, we show a comparison of the exact results and the ones obtained 
in first order perturbation theory, using the energy eigenvalues in Eqs.~(\ref{esph}) 
and (\ref{edef}), and the correspondence 
between the covariances and distribution of ground state probabilities of 
Eq.~(\ref{cov}). The results obtained in perturbation theory are remarkably close to 
the exact ones. There are some slight deviations for the spherical solutions, whereas 
in the deformed region there is an exact correspondence for the distribution of ground 
state angular momenta between the diagonal approximation and the full calculation. 

\begin{table}[t]
\caption{Distribution of ground state probabilities $P_{v}$ (in \%) for $N=15$ and 
$N=16$ bosons, obtained exactly (Full) and in the diagonal approximation (Diag) using 
Eq.~(\ref{esph}) for the spherical solution and (\ref{edef}) for the deformed case.}
\label{sphdef1}
\begin{tabular}{cc|rr|rr|rr}
& & \multicolumn{2}{|c|}{Spherical} & \multicolumn{2}{|c|}{Deformed} 
& \multicolumn{2}{|c}{Total} \tabularnewline
$N$ & $v$ & Full & Diag & Full & Diag & Full & Diag \tabularnewline
\hline 
15 & $0$ & 50.0 & 48.6 & 50.0 & 50.0 & 50.0 & 49.0 \tabularnewline 
   & $1$ & 33.3 & 34.7 &  0.0 &  0.0 & 25.0 & 26.0 \tabularnewline
   & $N$ & 16.7 & 16.7 & 50.0 & 50.0 & 25.0 & 25.0 \tabularnewline
\hline
16 & $0$ & 83.3 & 83.3 & 50.0 & 50.0 & 75.0 & 75.0 \tabularnewline 
   & $1$ &  0.0 &  0.0 &  0.0 &  0.0 &  0.0 &  0.0 \tabularnewline
   & $N$ & 16.7 & 16.7 & 50.0 & 50.0 & 25.0 & 25.0 \tabularnewline
\end{tabular}
\end{table}

The Hamiltonian of Eq.~(\ref{hibm1}) is a special choice, since in addition 
to the number of bosons and the angular momentum, also the boson seniority 
$v$ is a conserved quantum number. Let us now consider a Hamiltonian in which 
this is not the case. As a second example, we take a schematic IBM Hamiltonian 
that describes the transition between vibrational and rotational nuclei.  
In the notation of \cite{DS}, this Hamiltonian is given by 
\ba
H_2 &=& \cos \chi \, d^{\dagger} \cdot \tilde{d} +
\frac{\sin \chi}{N-1} \left[
( 2 \, s^{\dagger} s^{\dagger} - d^{\dagger} \cdot d^{\dagger} ) \, 
( {\rm h.c.} ) \right. 
\nonumber\\
&& \left. + ( 2 \, s^{\dagger} d^{\dagger} \pm \sqrt{7} \, 
d^{\dagger} d^{\dagger} )^{(2)} \cdot 
( {\rm h.c.} ) \right] ~.
\label{hibm2}
\ea
The Hamiltonian of Eq.~(\ref{hibm2}) has a spherical minimum for 
$-\pi < \chi < \arctan \frac{1}{9}$ and a deformed minimum for 
$\arctan \frac{1}{9}< \chi \le \pi$. For $\chi = \arctan \frac{1}{9}$, 
the system exhibits a first-order phase transition between spherical 
and deformed nuclei \cite{DSI,DS}. Just as in the previous example, the 
distribution of ground state angular momenta can be obtained exactly 
in a mean-field study. For $N=3k$ ($k=1,2,\ldots$) the ground state has 
$L=0$ in 75 \% of the cases and $L=2N$ (the maximum value of the angular 
momentum) in the remaining 25 \%. For all other values of the total number 
of bosons $N$, the ground state has either $L=0$, $L=2$ or $L=2N$ for 50 \%, 
25 \% and 25 \% of the cases, respectively. 

\begin{figure}[t]
\centering
\includegraphics[width=\columnwidth]{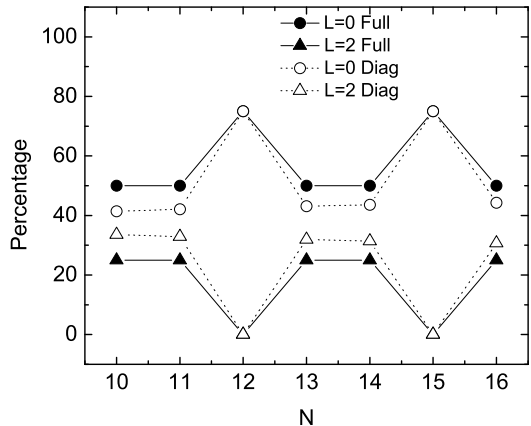}
\caption[]{(Color online) Percentage of ground states with angular momentum $L=0$ (circles), 
and $L=2$ (triangles) for $10 \le N \le 16$ bosons interacting via the Hamiltonian 
of Eq.~(\ref{hibm2}) with $-\pi < \chi \le \pi$ calculated exactly (solid lines) 
and in first order perturbation theory (dotted lines).}
\label{sphdef2}
\end{figure}

In order to test the correspondence between the covariances and the 
distribution of ground state angular momenta, we treat the Hamiltonian of 
Eq.~(\ref{hibm2}) in first order perturbation theory, in which the first 
term is treated as a perturbation to the second in the deformed region and 
{\it vice versa} in the spherical region. In Fig.~\ref{sphdef2}, we show 
a comparison of the percentages of ground states with $L=0$ and $L=2$ 
obtained exactly and in first order perturbation theory. Also in this case, 
the agreement is very good, although not at the level of precision as 
shown in Table~\ref{sphdef1} for the Hamiltonian of Eq.~(\ref{hibm1}).   
The deviations observed for the probabilities for ground states with $L=0$ 
and $L=2$ may be reduced in second order perturbation theory. 
The percentage of ground states with the maximum value of the angular momentum 
$L=2N$ is 25 \% in all cases, both exactly and in perturbation theory. 

In both examples, the distribution of the ground state angular momenta is 
reproduced to a great level of precision with the present procedure, which 
is a combination of perturbation theory to get energy eigenvalues which are 
linear in the random two-body matrix elements, and the relation that we 
derived between the covariances and the probabilities of the ground state 
angular momenta.    

The energies of many-body quantum systems are strongly correlated for the simple 
reason that they are all eigenvalues of the one and the same Hamiltonian. These 
strong correlations may modify the distribution of ground states from naive expectations, 
as was the case for the preponderance of $L=0$ ground states observed for calculations 
in the nuclear shell model and the IBM with random interactions \cite{JBD,BF}. 
Due to the strong correlations, it should not be surprising that the probability 
distribution of the ground state angular momenta for an ensemble of random two-body 
interactions is related to the covariances between energy eigenvalues. 

\section{Summary and conclusions}

In this manuscript, we addressed the problem of the emergence of regular features in 
many-body quantum systems with random interactions. As an example, we focussed on 
the probability distribution of ground state angular momenta in nuclear models, 
especially the unexpected observed dominance of $L=0$ ground states \cite{JBD,BF}. 
Even though many authors have tackled the subject, no general explanation for the 
preponderance of $L=0$ ground states has been found. A general feature of these systems 
is that the energy eigenvalues are strongly correlated, since all many-body matrix elements 
are expressed in terms of a relatively small number of random two-body matrix elements. 

In this article, we proposed to explore the correlations between energy eigenvalues 
(or covariances) in more detail. We established, to the best of our knowledge for the 
first time, an explicit relation between the probability distribution of ground state 
angular momenta and the covariances between energy eigenvalues. For diagonal (but in no 
way trivial) Hamiltonians, our formulas in terms of covariance coefficients are exact. 
For non-diagonal Hamiltonians it is not possible to give closed expressions but, by means 
of perturbation theory, nearly exact results can be found for the ground-state percentages. 

This new relation provides a purely statistical interpretation of the abundances 
of ground states with $L=0$ observed in numerical studies of many-body systems with 
random interactions which is, in contrast to all other explanations, valid for both 
fermions and bosons ({\it e.g.} the nuclear shell model and the IBM in nuclear physics). 
As an example of this procedure we studied two schematic nondiagonal Hamiltonians of 
the IBM and found an excellent agreement between the approximate results obtained in 
perturbation theory and the exact ones. Although these results were illustrated in the 
context of a specific model, we believe them to be sufficiently general to propose 
an entirely statistical explanation which is valid in general for any many-body quantum 
system. In future work, we will apply the present method to problems in the nuclear 
shell model. 

As a final comment, an explanation for the emergence of regular features in randomly 
interacting many-body systems which are based entirely on statistical arguments 
(covariances), may also provide a link to other unexpected results in statistical 
problems related to correlated data, such as Benford's and Zipf's laws.

\section*{Acknowledgments}

This work was supported in part by the grant IN113808 of PAPIIT, UNAM, Mexico.


\begin{thebibliography}{99}

\bibitem{Talmi}
See {\it e.g.} F. Iachello and I. Talmi, 
Rev. Mod. Phys. {\bf 59}, 339 (1987).

\bibitem{JBD}
C.W. Johnson, G.F. Bertsch, D.J. Dean, 
Phys. Rev. Lett. {\bf 80}, 2749 (1998).

\bibitem{BF}
R. Bijker and A. Frank, Phys. Rev. Lett. {\bf 84}, 420 (2000).

\bibitem{electronic}
Y. Alhassid, Rev. Mod. Phys. {\bf 80}, 2749 (1998).

\bibitem{benford}
J.W. Wittaker, SIAM, J. Appl. Math {\bf 43}, 257 (1983).

\bibitem{NPN}
R. Bijker and A. Frank,
Nuclear Physics News, Vol. 11, No. 4, 15 (2001) [arXiv:nucl-th/0111010]. 

\bibitem{msu}
V. Zelevinsky and A. Volya, Phys. Rep. {\bf 391}, 311 (2004).

\bibitem{zhao}
Y.M. Zhao, A. Arima and N. Yoshinaga, Phys. Rep. {\bf 400}, 1 (2004).

\bibitem{thomas}
T. Papenbrock and H.A. Weidenm\"uller, Rev. Mod. Phys. {\bf 79}, 997 (2007). 

\bibitem{BFP}
R. Bijker, A. Frank and S. Pittel, 
Phys. Rev. C {\bf 60}, 021302(R) (1999).

\bibitem{DD}
D. Dean, Nucl. Phys. A {\bf 682}, 194c (2001).

\bibitem{ymzhao}
Y.M. Zhao, A. Arima and N. Yoshinaga, Phys. Rev. C {\bf 66}, 034302 (2002).

\bibitem{BF2}
R. Bijker and A. Frank,
Phys. Rev. C {\bf 62}, 014303 (2000).

\bibitem{johnson}
C.W. Johnson and H.A. Nam, 
Phys. Rev. C {\bf 75}, 047305 (2007).

\bibitem{JBDT}
C.W. Johnson, G.F. Bertsch, D.J. Dean and I. Talmi,
Phys. Rev. C {\bf 61}, 014311 (1999).

\bibitem{mulhall}
D. Mulhall, A. Volya and V. Zelevinsky, Phys. Rev. Lett. {\bf 85}, 4016 (2000).

\bibitem{mf}
R. Bijker and A. Frank, Phys. Rev. C {\bf 64}, 061303(R) (2001);\\
R. Bijker and A. Frank, Phys. Rev. C {\bf 65}, 044316 (2002).

\bibitem{papenbrock}
T. Papenbrock and H.A. Weidenmuller, 
Phys. Rev. Lett. {\bf 93}, 132503 (2004); 
Phys. Rev. C {\bf 73}, 014311 (2006). 

\bibitem{chau}
P.C. Huu-Tau, A. Frank, N.A. Smirnova and P. Van Isacker, 
Phys. Rev. C {\bf 66}, 061302(R) (2002).

\bibitem{ibm}
F. Iachello and A. Arima, {\it The interacting boson model} 
(Cambridge University Press, 1987).

\bibitem{DSI}
A.E.L. Dieperink, O. Scholten and F. Iachello, 
Phys. Rev. Lett. {\bf 44}, 1747 (1980).

\bibitem{mex}
O. Casta\~nos, E. Chac\'on, A. Frank and M. Moshinsky, 
J. Math. Phys. {\bf 20}, 35 (1979);\\ 
A. Arima and F. Iachello, Ann. Phys. (N.Y.) {\bf 123}, 468 (1979).

\bibitem{DS}
A.E.L. Dieperink and O. Scholten, 
Nucl. Phys. A {\bf 346}, 125 (1980).

\end{thebibliography}
\end{document}